\documentclass[reprint,amsmath,amssymb, aps, superscriptaddress]{revtex4-1}

\usepackage{graphicx}
\usepackage{dcolumn}
\usepackage{bm}
\usepackage{indentfirst}
\usepackage{natbib}
\usepackage{hyperref}
\begin{document}
	
	\title{Design of Magneto-Optical Traps for Additive Manufacture by 3D Printing}
	
	\author{Yijia Zhou}
	\affiliation{School of Physics and Astronomy, University of Nottingham, Nottingham, NG7 2RD, UK}
	\affiliation{Department of Physics, Fudan University, Shanghai, 200433, P.R.China}
	\author{Nathan Welch}
	\affiliation{School of Physics and Astronomy, University of Nottingham, Nottingham, NG7 2RD, UK}
	\author{Rosemary Crawford}
	\affiliation{School of Physics and Astronomy, University of Nottingham, Nottingham, NG7 2RD, UK}
	\author{Fedja Oru\v{c}evi\'{c}}
	\affiliation{School of Physics and Astronomy, University of Nottingham, Nottingham, NG7 2RD, UK}
	\affiliation{Present address: Department of Physics and Astronomy, University of Sussex, Falmer Campus, Brighton, BN1 9QH, UK}
	\author{Feiran Wang}
	\affiliation{School of Physics and Astronomy, University of Nottingham, Nottingham, NG7 2RD, UK}
	\author{Peter Kr\"{u}ger}
	\affiliation{School of Physics and Astronomy, University of Nottingham, Nottingham, NG7 2RD, UK}
	\affiliation{Present address: Department of Physics and Astronomy, University of Sussex, Falmer Campus, Brighton, BN1 9QH, UK}
	\author{Ricky Wildman}
	\affiliation{Faculty of Engineering, University of Nottingham, Nottingham, NG7 2RD, UK}
	\author{Christopher Tuck}
	\affiliation{Faculty of Engineering, University of Nottingham, Nottingham, NG7 2RD, UK}
	\author{T. Mark Fromhold \thanks{mark.fromhold@nottingham.ac.uk}}
	\affiliation{School of Physics and Astronomy, University of Nottingham, Nottingham, NG7 2RD, UK}

	\begin{abstract}
		A key element in the study of cold atoms, and their use in emerging quantum technologies, is trapping the atoms in an ultra-high vacuum (UHV) chamber. Many methods have been used to trap atoms including atom chips and magneto-optical traps (MOTs). However, the bulky apparatus, and current-carrying coils, used so far in most MOTs restrict the reduction of power and physical size, as required for quantum technology applications. The advent of 3D printing technology now offers a new route to making MOTs with current paths that can be freely shaped and shrunk to several centimetres, thereby helping to reduce the power consumption and simplify the production of the MOT itself. In this paper, we present designs for 3D printed MOTs and analyse their performance by using COMSOL simulations. We predict that the 3D-printed conductors can create magnetic fields with gradients around 15 G/cm and passing through zero, as required for atom trapping, with Joule heating as low as 0.2 W.
	\end{abstract}
	
	\maketitle
	
	\section{Introduction}
	Many devices have been developed to trap neutral atoms in an ultra-high vacuum (UHV) environment by using light, magnetic and electric fields. One of these, the magneto-optical trap (MOT), comprises laser light and a magnetic field \cite{MOT1,MOT2,MOT3,MOT4, chipreview} and operates by exploiting the atom-photon interaction, which depends on both the position and momentum of the atoms. Atoms can be excited by absorbing a photon and then spontaneously emitting another. The emission is usually isotropic, and thus does not change the mean momentum of the atom cloud. However, the absorption process is anisotropic and can thus slow the atoms if the frequency and polarization of the light is chosen correctly. The external magnetic field controls the internal states of the atoms and the absorption rate. For the MOT to work, we need the magnitude of the magnetic field to increase with increasing distance from the trap centre. \par
	The traditional experimental MOT set-up includes anti-Helmholtz coils outside the UHV chamber. The magnetic field is highly inhomogeneous and, in each direction, its strength varies approximately linearly around the trap centre. Another well-known trap configuration is the Ioffe-Pritchard trap comprising four horizontal wires \cite{Ioffe1, Ioffe2}, which provide a quadrupole magnetic field in the vertical plane, and two vertical rings carrying parallel currents, which trap the atoms horizontally. \par
	Traditional large (10s of cm) laboratory-based MOTs often have hand-made coils positioned outside the UHV chamber. It is laborious to make such coils, which inhibit the miniaturisation of the apparatus and consume high power of $\sim$ 20 W. \par
	In this paper, we highlight the potential of 3D printing technology \cite{3dprint}, to make current-carrying components for the MOTs, whose geometries can be shaped freely in order to reduce power and fit with surrounding components. Due to the scalable nature of such Additive Manufacturing techniques, from metre down to sub-micron scales, they may also be transferable to atom chip structures \cite{chip1,chip2,chip3}. We present three designs for MOTs suitable for production via 3D printing. We start from curved and twisting metal bars, which illustrate the principle of shaping the current paths in order to produce the required magnetic field. Guided by this, we then consider designs with successively more compact structures and volume-filling conductors designed to achieve lower power dissipation and more accurate field landscapes. For most MOTs, the magnetic field gradients along the $x,y$ and $z$ directions are chosen to be in the ratio of approximately 1:1:-2 in order to produce an atom cloud with similar dimensions in all three directions. The most efficient gradients for Rb-87 cooling are around 15 G/cm \cite{Rbtrap1, Rbtrap2} and in our design we obtain 9.0, 9.2 and -17.6 G/cm respectively. A major advantage of the 3D-printed traps is that we can reduce their size to be the same order as the laser beam diameter, thereby enabling the current to flow nearer to the trapped atom cloud. Thus, the heat dissipation in the conductors can be reduced to 0.2 W in our simulations. As noted above, such miniaturization and low power are essential for the development of commercial applications in quantum technologies such as sensors and clocks.
	
	\section{Principles of Magneto-optical Trap Operation}
	\begin{figure}
		\centering
		\includegraphics[width=8.5cm]{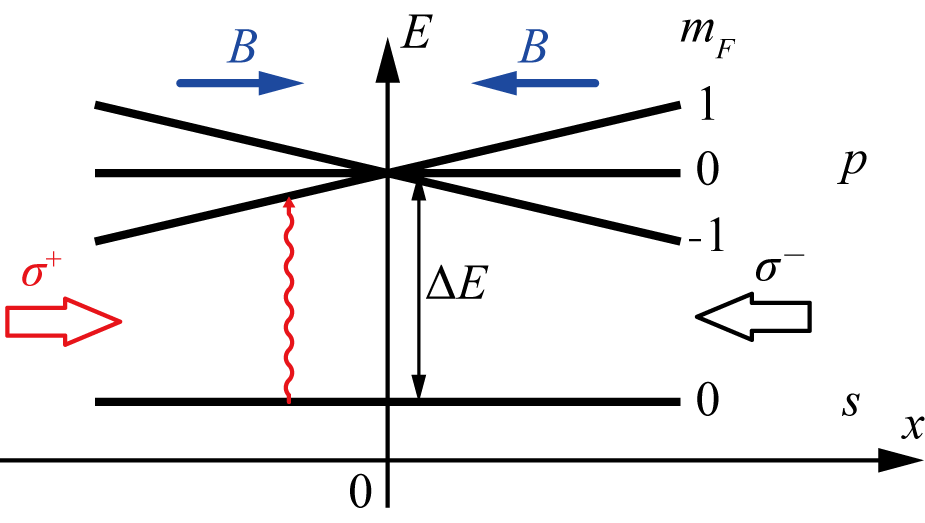}
		\caption{\label{basic} Schematic diagram showing the operation of a MOT for atomic motion along the $x$-axis (a similar picture holds along the $y$ and $z$ directions): The strength of the magnetic field is proportional to $x$, and its direction is shown by the blue arrows. The Zeeman effect splits the energy of atoms by an amount proportional to $m_F$ (values shown right). When an $s$-state atom moves from $x=0$ to the left, and absorbs a photon (represented by the red arrows) with energy slightly below $\Delta E$, it will be excited to the $m_F=1$ state. To maintain momentum conservation, the atom is slowed by photon absorption. The de-excitation process emits photons randomly in all directions so that, on average, the atom slows. A similar deceleration occur for right-moving atoms that absorb left-moving photons (black arrow labelled $\sigma^-$). }
	\end{figure}
	MOTs enable many atoms, between $10^7\sim 10^{10}$, to be cooled to $\mu$K temperatures. A unique feature of their operation is that they can selectively cool atoms according to their position. When the atoms are not at the trap centre, the light beams provide a force that push the atoms towards the centre. In the simplest model for this restoring force, the atoms are taken to have only two internal states: a ground $s$-state and an excited $p$-state. The $p$-state splits into three substates in a magnetic field via the Zeeman effect. In MOTs, the magnetic field strength is proportional to the distance along the $x,y$ and $z$-axes from the trap centre. The excited state energy consequently splits as the atoms move away from the trap centre, as shown in Fig.\ref{basic}. The frequency of the laser beams applied along the axes are red detuned to be slightly lower than the ground to excited state transition frequency. Due to the small energy difference between the Zeeman-split energy levels, absorption of photons preferentially occurs as the atoms move away from the trap centre, which slows the atoms. The excited atoms emit photons spontaneously. This emission is isotropic and therefore does not change the mean momentum of the atom cloud. For convenience we consider 6 laser beams for cooling, making use of the conservation of angular momentum in photon transitions. For example, suppose one atom moves leftwards, as shown in Fig.\ref{basic}, where $x<0$ and the magnetic field points to the right. If a red-detuned right-going photon has $\sigma^+$ polarization, it will excite the atom to the $m_F=1$ state. Left-going $\sigma^-$ photons are also red-detuned, but the frequency needed to excite the atom to the $m_F=-1$ state is higher, so such photons will be absorbed with much smaller probability. Thus one pair of laser beams can slow atoms in one direction. Three pairs of laser beams will slow and reduce the temperature of the atom cloud as a whole. For this mechanism to work, the current must produce a quadrupole magnetic field at the trap centre. \par 
	Traditionally, a pair of anti-Helmholtz coils generates the required field profile. Anti-Helmholtz coils comprise a pair of parallel coils with current flowing in opposite directions as shown in Fig.\ref{trap1}($a$)(left). The magnetic field generated by each coil cancels at the trap centre, whilst the field gradients add. Maxwell's equations require that $\partial B_x / \partial x + \partial B_y / \partial y + \partial B_z / \partial z = 0$ at the trap centre. For the particular symmetry of anti-Helmholtz coils, this relation yields $\partial B_x / \partial x = \partial B_y / \partial y = -(1/2) \partial B_z / \partial z$. \par
	The conventional way of manufacturing anti-Helmholtz coils is to use multiple turns of copper wires held outside the UHV chamber. This makes the coils far from the atom cloud and requires huge currents ($\sim$ 100 A) in the wires. To utilize the space between the coils, and thus achieve more power-efficient trapping, Ioffe and Pritchard independently designed a magnetic trap to generate the field whose $x,y$ components are quadrupole as shown in Fig.\ref{trap1}($a$)(right) \cite{Ioffe1,Ioffe2}. The four parallel wires generate the quadrupole magnetic field along the $x$ and $y$ directions, and an additional pair of coils provides the magnetic field for $z$-directional trapping. \par
	To further reduce the heat dissipation resulting from the currents, we have investigated compact conductor geometries, which wrap closely around the atomic cloud, and are suitable for manufacturing by using 3D printing technology. Our designs comprise either 2 or 4 separate conducting parts and 4 current terminals. We consider topologies inspired by both the Ioffe-Pritchard trap and anti-Helmholtz coils.
	
	\section{Twisted Metal Bars}
	\begin{figure}[h]
		\centering
		\includegraphics[width=\linewidth]{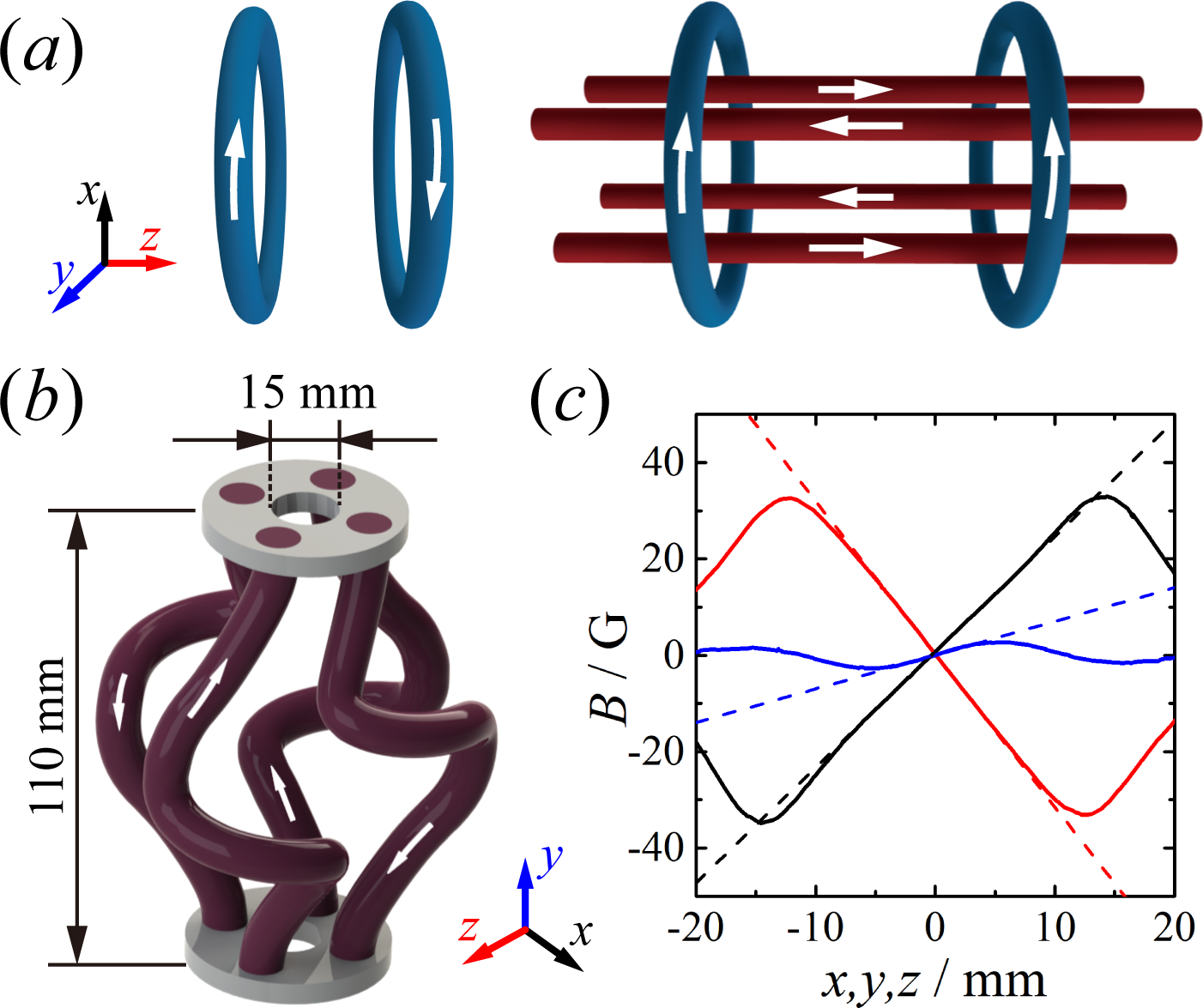}
		\caption{\label{trap1} (a) The traditional anti-Helmholtz coils (left) and Ioffe-Pritchard trap. The white arrows show the current directions. (b) A MOT design using twisted metal bars to achieve 3D trapping. The white discs are dielectric mounts and the purple parts are the bars. The white arrows show the current direction in each bar. (c) Calculation of the magnetic field magnitude, $B$, versus position along the $x,y$ and $z$ directions (black, blue and red curves respectively). The gradients near the trap centre along the $x,y$ and $z$ directions are 15.1, 6.5 and -22.1 G/cm (black, blue and red dashed lines).
		}
	\end{figure}
	To illustrate how MOTs can also be made by producing curvilinear current paths that are neither straight nor circular, we now consider the use of twisted metal bars, each with a circular cross-section, to guide the current and generate the magnetic field for the MOT. Inspired by the Ioffe-Pritchard trap, we first analyse a twisted cage consisting of four twisted metal bars shown in Fig.\ref{trap1}($b$). The shape of the bars is guided by parametric equations \cite{BarEq}. The current in adjacent bars flows in opposite directions (white arrows). The vertical current component contributes to the horizontal magnetic field, whilst the twisted current paths generate the vertical field. Together, the four bars make the field vanish at the trap centre and increase linearly around it. This scheme can provide the magnetic field required for a MOT. The height of the trap in Fig.\ref{trap1}($b$) is 110 mm, and the outer width 55 mm. The copper bars are 10 mm in diameter. The holes in the bottom and top insulating mounts (white), and the space between wires allow access of 15 mm diameter orthogonal laser beams. The trap is designed using 3D CAD software Solidworks and the physical fields are simulated by COMSOL Multiphysics. We calculated the current density and magnetic field profile with fixed voltages at the bar ends. The solution domain is a cube three times larger than the trap. In Fig.\ref{trap1}, we plot the field along three axes through the trap centre where the field gradients have approximately constant values of 15.1, 6.5 and -22.1 G/cm along the $x,y$ and $z$-axes respectively. \par
	The current needed to make the trap is 100 A and the total electric power dissipation via Joule heating is 14.9 W for copper. Although illustrative, this is currently not the best design we can produce. There is still a lot of unfilled space between the conductors and the trap centre, making the trap power consumption unecessarily high. We found that a high-performance MOT should follow several key design criteria: \par
	\begin{itemize}
		\item Conductors should wrap the optical molasses region tightly and fill as much space as possible around it;
		\item The trap conductors should be as compact as possible;
		\item Each bar/conducting volume should always wind in the same direction either clockwise or anticlockwise to maximize its contribution to the magnetic field.
	\end{itemize} \par
	The twisted bar design in Fig.\ref{trap1}($b$) does not satisfy all of the above criteria. It is large and the small-area twisted conductors are always winding in the same direction, which decreases the performance of the device. In addition, the field gradient ratios are slightly different from those, 1:1:-2, typically used in MOTs and the current required to generate these gradients is high. Consequently, in the next section we consider a structure that also employs curvilinear current flows to generate the trapping fields, but has improved field and power performance.
	
	\section{Compact Conductor Design}
	\begin{figure}[h]
			\centering
			\includegraphics[width=\linewidth]{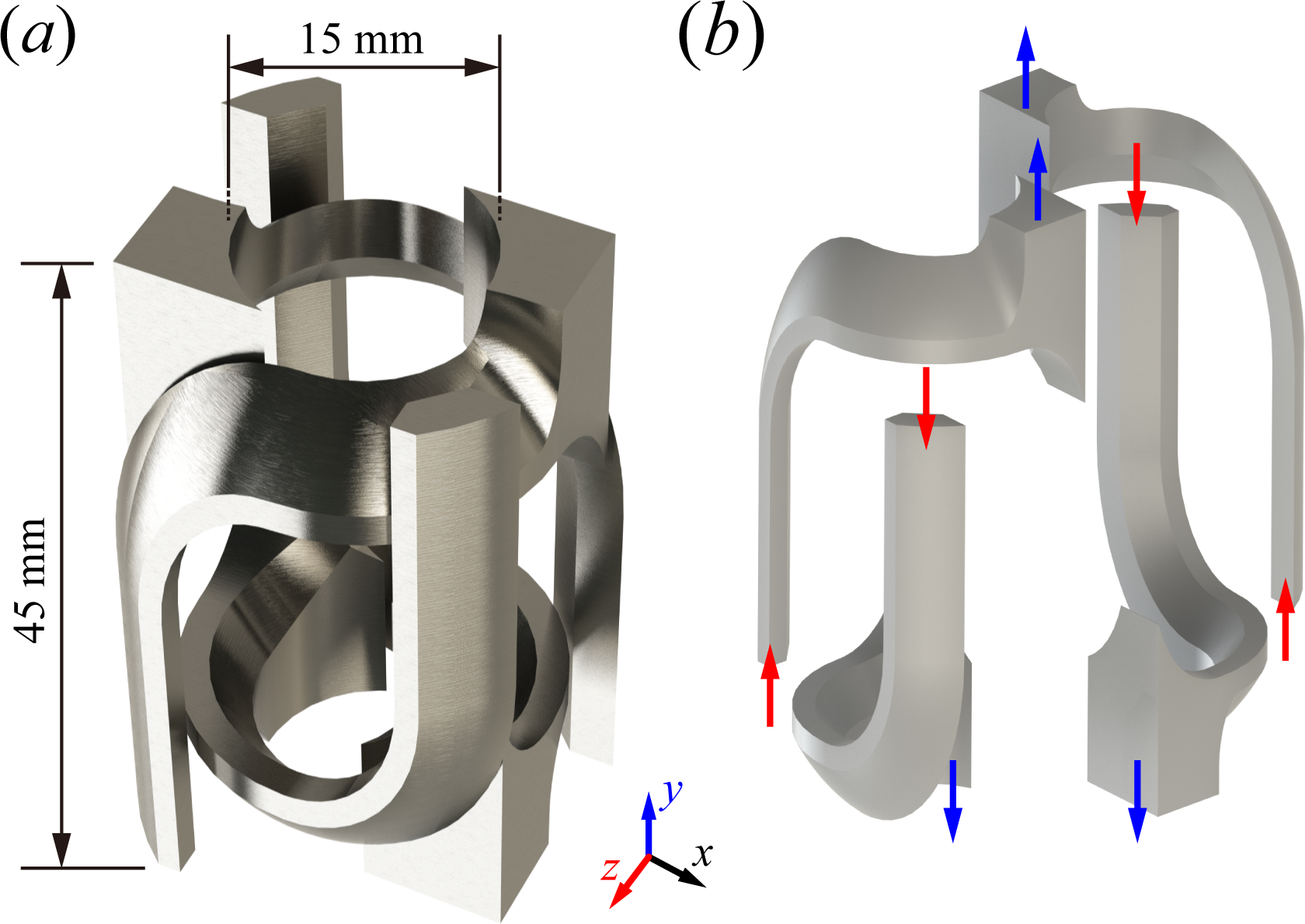}
			\caption{\label{trap2} (a) Design for a 3D-printed version of the current topology shown in Fig.\ref{trap1}(b). The four conducting parts are electrically isolated, and the gaps between them can be filled with dielectric (not shown). (b) An exploded view of the trap. Red and blue arrows indicate the current input and outflow.
			}
	\end{figure}
	We now present an improved conductor design, shown in Fig.\ref{trap2}, which maintains the topology of the current paths shown in Fig.\ref{trap1} but offers superior performance. In this case, the total height of the trap is 45 mm, and its outer width is 24 mm. The holes for the laser beam access are also 15 mm in diameter. The minimum gap between the four conductors is 0.5 mm, which can be adjusted according to the material stability in vacuum. The design requires four pairs of voltage contacts to run the trap and control the current in each conducting part. Compared to the twisted bar design in Fig.\ref{trap1}($b$), this structure better satisfies the three design criteria specified in Section III. It is smaller, makes better use of space around the atom cloud, and all the current flows contribute positively to the desired field profile. \par
	The current in the conductors can be divided approximately into vertical and horizontal components, with the former creating the magnetic field in the $xz$-plane and the latter generating the field along the  $z$-axis. The currents in adjacent conductors are anti-parallel at the contacts and, at the trap centre, provide a linear magnetic gradient along the $x,y$ and $z$-axes. To change the gradient ratio in the three directions, one may easily adjust the length of the conducting prongs. To produce the field gradients close to the 1:1:-2 ratio typically used for MOTs, we choose the height of the device to be larger than its width. To further reduce the resistance and power consumption, one may further increase the thickness of the volumetric conductors. \par
\begin{figure}
		\centering
		\includegraphics[width=\linewidth]{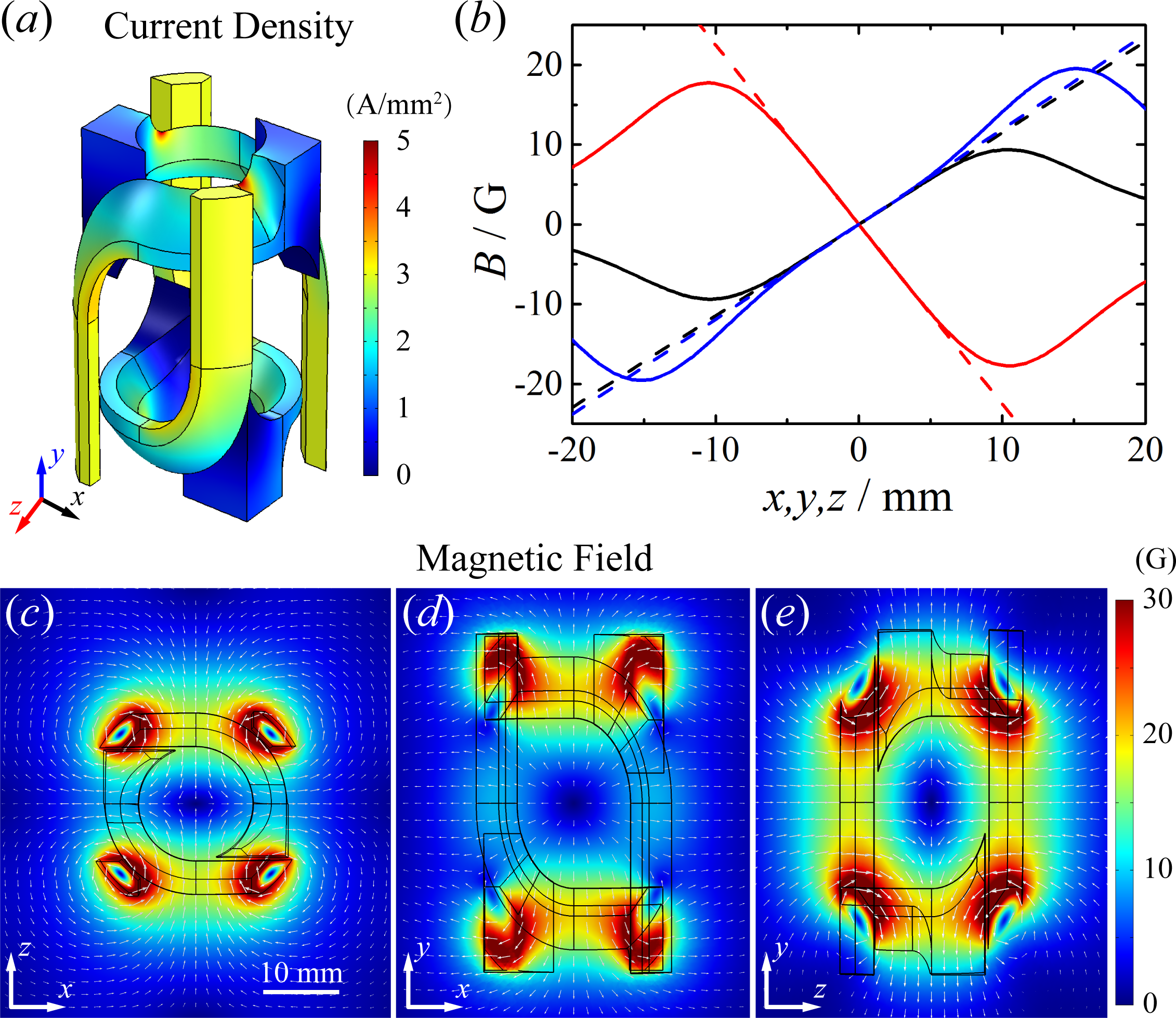}
		\caption{\label{field2} (a) Current density distribution calculated for the MOT shown in Fig.\ref{trap2}, with each conductor carrying 40 A. (b) Calculation of the magnetic field magnitude versus position along the $x,y$ and $z$-axes (black, blue and red solid lines). The gradients along the $x,y$ and $z$-axes at the trap centre are 11.5, 11.9 and -22.5 G/cm respectively (black, blue and red dashed lines). (c-e) The magnetic field magnitude in the $xz,xy$ and $zy$-planes through the centre of the MOT. In each plane, there is a minimum at the centre. The gaps between the four conductors are just large enough for a 1.5 cm diameter laser beam to pass through.
		}
	\end{figure}

	We designed the trap in Solidworks and simulated it in COMSOL. In Fig.\ref{field2}($a$), we show a colour map of the current density on the surface of the conductors. In Fig.\ref{field2}(b), we show the corresponding magnetic field components $B_x$, $B_y$, $B_z$ along the $x,y$ and $z$-axes and linear fits at the trap centre. The field variations across the $xz,xy$ and $zy$-planes through the trap centre are shown in Fig.\ref{field2}(c-e), where the colour scale shows the field magnitude. \par
	In general, the current density is highest at sharp bends in the conductors. To avoid current hot spots, we therefore fill the bend regions with curving surfaces, which is possible using 3D printing. The current on the inner conductor surfaces is higher than at the outer ones, which helps to increase the magnetic field in the trap, thus reducing the current and power consumption required. In Fig.\ref{field2}(b), the linear fitted slopes near the centre along the $x,y$ and $z$-axes are 11.5, 11.9 and -22.5 G/cm respectively. The total Joule heat in the trap is 0.4 W for copper conductors. \par
	A shortcoming of this design is that we need to print four parts (Fig.\ref{trap2}) and either fix them separately, or print dielectric material between each part to maintain the integrity of the structure. Such an arrangement is possible, but still a challenge for current Additive Manufacturing techniques. In the next section, we consider further simplification of the design to remove this issue by reducing the number of separate conductors.

	\section{Two-Piece Conductor Design}
	\begin{figure}[h]
		\centering
		\includegraphics[width=8.5cm]{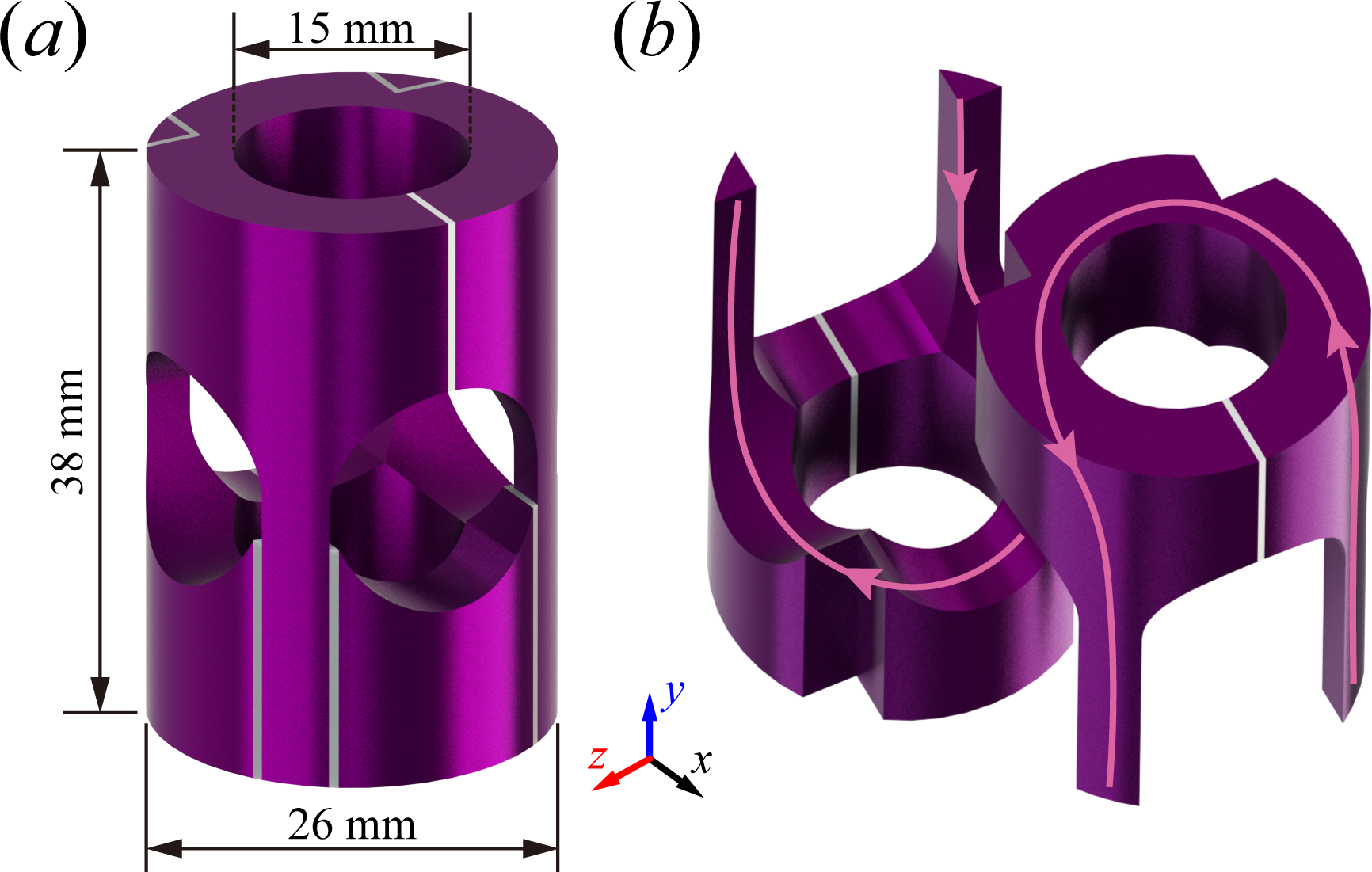}
		\caption{\label{trap3} (a)A 3D-printed MOT design with conductors that have a cylindrical outer curved surface. The shape is inspired by the Ioffe-Pritchard trap. (b) An exploded view of this model showing the two conductors. The pink lines and arrows show the directions of the current. Gaps between the conductors appear white.
			}
	\end{figure}
	Suppose that we produce three orthogonal laser-access holes, with circular cross sections, in a cylindrical conductor and then separate that conductor into two nesting parts as shown in Fig.\ref{trap3}. The current will flow in and out of the straight arms through a roughly circular path in between. The parallel arms will provide a quadrupole field in the trap centre and the approximately circular current will create the field pointing towards the trap centre along the axial directions. This geometry takes advantage of filling almost the whole volume surrounding the laser beams with conductor, thereby reducing the resistance of the current paths. The ratio of the thickness of the arms of the conductor to its outer ring diameter is important to ensure that the gradients are constant and close to the ratio 1:1:-2; typically we achieved 1.0:1.0:-1.9. Our design in Fig.\ref{trap3} permits the required magnetic field to be generated with typically 25 A in each conductor. \par
	The trap in Fig.\ref{trap3} is 38 mm tall, and the outer diameter is 26 mm. The arms are 3.1 mm wide at their narrowest point, i.e., the upper and lower prongs in Fig.\ref{trap3}($b$). The three holes for laser beams are 15 mm in diameter. The dimensions are optimized with respect to power consumption. The white stripes in Fig.\ref{trap3} are dielectric material to insulate the two conductors, or simply a vacuum gap if the positioning and clamping of the conductors is accurate enough. Currently, we choose a gap of 0.5 mm. Despite possible capacitance effects, the gap should be as thin as possible to minimize resistance and power dissipation. The approximate current paths are shown by the pink arrowed curves in Fig.\ref{trap3}($b$). Note that these have the same overall topology as the current paths through the four conductors shown in Fig.\ref{trap2}.\par
	\begin{figure}
		\centering
		\includegraphics[width=\linewidth]{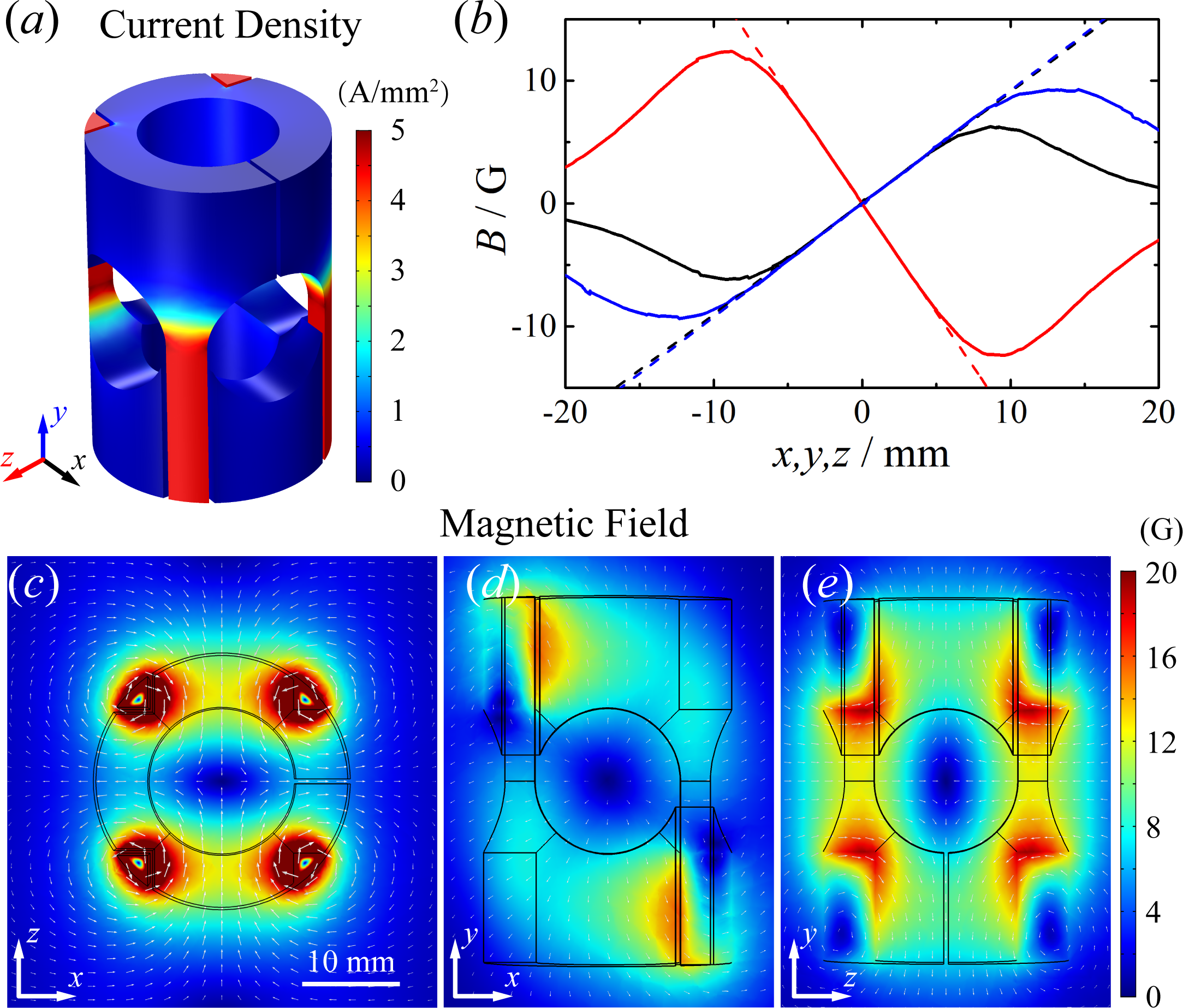}
		\caption{\label{field3} (a) Current density distribution calculated for the MOT shown in Fig.\ref{trap3}, with each conductor carrying 25 A in total. (b) Calculation of the magnetic field magnitude versus position along the $x,y$ and $z$-axes (black, blue and red solid lines). The gradients along the $x,y$ and $z$-axes at the trap centre are 8.98, 9.20 and -17.6 G/cm respectively (black, blue and red dashed lines). (c-e) The magnetic field magnitude in the $xz$,$xy$ and $zy$-planes through the centre of the MOT. In each plane, there is a minimum at the centre. The circular holes in the trap are just large enough for a 1.5 cm diameter laser beam to pass through.
		}
	\end{figure}
	The current density in each arm is $\sim$ 5 A/mm$^2$. In the ring-like regions, by contrast, the current density is significantly smaller, below 1 A/mm$^2$. By design, there is no hot spot in the current density so the trap will be heated uniformly. Our simulations show that the contributions to the magnetic field from the currents in the arms and rings are actually equivalent, which leads to the ratio of the gradient along the $y$ and $z$ directions being the same. \par
	The trap is designed using Solidworks and the field is calculated with COMSOL. We set the current to be 25 A in each conductor, and calculate the static magnetic field within the interior space. Fig.\ref{field3}($a$) shows a colour map of the current density on the surface of the conductors. Fig.\ref{field3}($b$) shows the magnetic field components $B_x$, $B_y$, $B_z$ along the $x,y$ and $z$-axes and linear fits (dashed lines) near the trap centre. The field gradients along the $x,y$ and $z$-axes are 9.0, 9.2 and -17.6 G/cm respectively with an uncertainty of $\sim$ 0.03 G/cm in each case. The total Joule heat in the trap is $\sim$ 0.2 W for copper. However, the most common materials for 3D printing are titanium and nickel, which have $\sim$ 10 times larger resistance than copper. Thus to provide the same magnetic field using these materials will require about 2 W in power dissipation in the conductors. \par
	To keep the ratio of the field gradients to be 1:1:-2 along the three directions, a trap of the form in Fig.\ref{trap3} is close to ideal. Elongating or thickening the trap will allow this ratio to be tuned as required for the other MOT geometries. \par
	MOTs usually work under UHV conditions, and the trapping process is almost adiabatic. Consequently, heat can only be dissipated through the ends of the arms. The heat transfer is proportional to the transfer area and temperature difference. Considering the total power dissipation to be $\sim$ 0.2 W for copper, and the transfer area to be $\sim 4\times10^{-5}\rm{m^2}$, in order to maintain the temperature increase below an experimentally-tolerable limit of 200 K, the heat transfer coefficient should be around 25 $\rm{W/m^2K}$. This is practical for experimental set-ups and, if additional external cooling can be applied to the trap, the operating temperature can be made lower than 470 K.
	
	\section{Conductor Parameters and Optimization}
	The size and density of the trapped atom clouds depend strongly on the magnetic field and trapping efficiency. Suppose that the trap volume is $V_T=l^3$, where $l$ is an effective linear size, the cross sectional area of the laser beams is $A$, the resistance of the conductor is $R$, the operating temperature of the conductors is $T$, and the heat dissipation rate is $\rho$. Then we can consider the general scaling relations: $V_T \propto A \times l \propto l^3$, $R \propto  l / A \propto l^{-1}$, $\rho \propto A \times T \propto l^2$. To avoid the traps becoming too hot, the radiation rate should equal the electric power, i.e., $I^2R \propto A \propto l^{2}$. So the current passing through the trap must follow $I \propto l^{3/2}$. The Biot-Savart law tells us that the magnetic field $B \propto I/l^2 \propto l^{-1/2}$, and the gradient $\nabla B \propto l^{-3/2}$. Therefore, to reduce the operating power, it is advantageous to make the trap smaller providing that all other experimental requirements can still be satisfied. \par
	Reduce the power consumption can also be achieved by thickening the conductors and reducing their resistance. However, if the conductors are too thick, the current will be far from the trap centre and generate a lower magnetic field. In practice, we need the current to be as small as possible to minimize the power. As there is, in effect, only one current turn within the bulk conductors that we consider, and the current needs to be as large as 25 to 100 A, a special power supply is needed. Another key consideration is the length of the trap, which should be neither too long nor too short. As mentioned above, the ring-like regions of the conductors are responsible for generating the field variation along the axial direction, which should be comparable to that in the $x$ and $y$ directions, as is the case for our model designs. We may be able to further improve the geometry of the conductors and reduce the difference between the current density in the arms and the ring-like regions in Fig.\ref{field3}. We can also remove some redundant parts of the conductors in order to fit them more tightly around real experimental set-ups and components of, for example, quantum sensors.
	
	\section{Conclusion}
	Compared to traditional MOT traps, the designs show good performance, especially the device in Fig.\ref{trap3}. According to our simulations, we need a 25 A current source to drive the trap to generate the magnetic field profile needed for a MOT. The resulting power consumption in the conductors is merely 0.2 W for copper. Whether this idea works in practice depends on the capability of 3D printing technology, which, for metals, presently focuses on titanium, aluminum, etc., whose resistance is larger than copper. Whether the traps are stable, especially the dielectric layers between the conductive parts, also needs to be verified in experiment. The next step is to design and build a multi-turn 3D printed conductive geometry, which can help to control the spatial current density distribution in a more subtle way as well as reducing the current and power dissipation and improving  the fidelity of the magnetic field landscape. It may also be possible to use Additive Manufacturing to produce atom traps with more complex geometries, including multiple potential minima and lattices with sub-micron scales.
	
	\section*{Acknowledgements}
	This work is funded by the Engineering and Physical Sciences Research Council (EPSRC) through the UK Quantum Technology Hub for Sensors and Metrology. Y. Zhou is sponsored by the China Scholarship Council (CSC).


\begin{thebibliography}{10}
\bibitem{BarEq}
		In spherical coordinates $\{r, \theta, \phi\}$, for $0 < \theta < \pi/4$,
		$r(\theta) = l\big{[}2-\sqrt{1-(4\theta/\pi-1)^2}\big]/4$, $\phi(\theta) =
		\pi/\big[3+3e^{-10(\theta-\frac{\pi}{3})}\big]$; for $\pi/4 < \theta <
		3\pi/4$, $r(\theta) = l/4$, $\phi(\theta) =
		\pi/\big[3+3e^{-10(\theta-\frac{2\pi}{3})}\big]$; for $3\pi/4 < \theta <
		\pi$, $r(\theta) = l\big[2-\sqrt{1-(4\theta/\pi-3)^2}\big]/4$, $\phi(\theta)
		= \pi/\big[3+3e^{-10(\theta-\frac{\pi}{3})}\big]$. Here $l=110$ mm is the
		height of the trap.
\end{thebibliography}

\end{document}